\documentstyle[11pt,moriond,epsfig]{article}

\def\ra{\rightarrow}

\def\be{\begin{equation}}
\def\ee{\end{equation}}
\def\bea{\begin{eqnarray}}
\def\eea{\end{eqnarray}}

\begin{document}
%\vspace*{1cm}
\title{SUPERCURRENT NOISE IN SNS JUNCTIONS}

\author{D.V. AVERIN }

\address{Department of Physics and Astronomy,  SUNY 
at Stony Brook, Stony Brook NY 11794}

\maketitle\abstracts{Quasiclassical Green's functions of a 
short superconductor/normal metal/superconductor (SNS) junction 
are calculated for multi-mode junction with arbitrary electron 
transmission properties. They provide the basis for description 
of all of the junction transport characteristics, and are used to 
calculate the equilibrium spectral density of supercurrent 
fluctuations.}

\section{Introduction} 

The process of multiple Andreev reflections (MAR) is the 
dominant mechanism of electron transport in Josephson junctions 
with large electron transparency. One of the main qualitative 
features of MAR is the coherent transfer of charge by large 
quanta, the size of which increases with decreasing bias voltage 
$V$. At low bias voltages, $eV\ll \Delta$, where $\Delta$ is the 
superconducting energy gap, this leads to strong increase of 
non-equilibrium current noise in the MAR regime, the fact that 
presently attracts interest to the noise properties 
of high-transparency junction both in theory~\cite{b1,b2,b3} and 
experiment~\cite{b4,b5,b6}. At $V$ smaller than the quasiparticle 
energy relaxation rate $\gamma$, $eV\ll \hbar \gamma$, the MAR-noise 
saturates and turns into equilibrium supercurrent noise that is 
best described in terms of the thermal fluctuations of the 
occupation factors of the Andreev states carrying supercurrent. 
Since the dynamics of these occupation factors is driven by 
inelastic transitions, comprehensive theory of the current noise in 
high-transparency junctions should be based on the 
Green's functions formalism which provides a rigorous description 
of the inelastic processes. Until now, such a theory was developed 
only for the single-mode ballistic junctions~\cite{b1}. The aim of 
this work is to calculate quasiclassical Green's functions of a 
short multi-mode Josephson junction with arbitrary electron 
transmission properties. They are used then to derive the 
general expression for equilibrium supercurrent noise and to find 
this noise in disordered SNS junctions with diffusive electron 
transport.

\section{Junction model and Green's functions} 

The basic model of a short Josephson junction with large electron 
transparency is a constriction between two superconductors that 
supports $N$ propagating electron modes. If the constriction length 
$d$ is much smaller than the superconductor coherence length $\xi$, 
the only characteristic of the constriction relevant for electron 
transport is the elastic scattering matrix $S$: 
\be 
S= \left( \begin{array}{cc} r & t \\ t' & r'
\end{array} \right) \, , 
\label{1}  \ee
which is energy-independent in the relevant energy 
range set by the superconducting energy gap. Unitarity of $S$ 
implies that the $N\times N$ transmission and reflection matrices 
can be expressed as: $t=u_1^{\dagger}\sqrt{D} u_2\,$, $t'= 
u_3^{\dagger} \sqrt{D} u_4\,$, $r=u_1^{\dagger} \sqrt{R}u_4\, $, 
$r'= -u_3^{\dagger} \sqrt{R}u_2\,$, where $D$ is the diagonal matrix 
of $N$ transmission probabilities, $R=1-D$, and $u_j$ are unitary 
matrices.   

Transport characteristics of such junctions can be expressed in 
terms of their quasiclassical Green's functions. Calculation of 
these functions for the multi-mode junctions presented here 
follows closely similar calculation~\cite{b7} in the single-mode 
case. The quasiclassical Green's functions can be written as 
\be
G(z,z')=\sum_{\pm}(E_{\pm}e^{\pm ik_F(z-z')} +F_{\pm}e^{\pm 
ik_F(z+z')} )\, ,
\label{2} \ee
where $z,z'$ are the coordinates along the current flow, and $k_F$ 
is the Fermi momentum. The amplitudes $E$ and $F$ are 
slowly varying functions of coordinates in the junction electrodes 
and have matrix structure in the transverse mode space. Because 
of the electron  scattering, they change rapidly in the junction 
region ($z\simeq 0$). Since the junction is short, $d\ll \xi$, 
rapid variation of the amplitudes $E$ and $F$ across it can be 
described as the boundary condition relating them at $z\ra -0$ 
and $z\ra+0$. This condition is derived from the fact that in the 
junction region, 
$G(z,z')$ satisfy regular Schr\"{o}dinger equation as a function 
of $z$ and $z'$, and therefore can be expressed in terms of the 
products of the two scattering states, 
\be
G(z,z')=\sum_{i,j=1,2}\psi_i(z)\lambda_{ij}\psi_j(z')\, . 
\label{n1} \ee 
Here $\psi_1$ is the solution of the Schr\"{o}dinger equation for
particles incident from the left, 
\be
\psi_i(z) =  \left\{ \begin{array}{l} e^{ik_Fz}+ r e^{-ik_Fz}\, , 
\;\; z\ra -\infty \, ,\\  
t e^{ik_Fz} \, , \;\; z\ra \infty \, , \end{array} \right. 
\label{n2} \ee 
and $\psi_2$ is similar solution for scattering from the right. 
Note that ``infinity'' in eq.\ (\ref{n2}) means $\mid \! z\!\mid \gg 
d$ and still corresponds to $z\simeq 0$, i.e. $\mid \! z\! \mid\ll \xi$, 
on the length scale of variations of the amplitudes $E$ and $F$ in the 
electrodes that is given by the coherence length $\xi$. Comparison of 
expansion of $G(z,z')$ obtained from eqs.\ (\ref{n1}), (\ref{n2}), 
and similar equation for $\psi_2$, in the region to the left of the 
junction with the expansion to the right establishes four relations  
between the amplitudes $E$ and $F$ in these two regions: 
\bea  
t'^{\dagger}(F_+(+0)-E_+(+0)r')=(F_+(-0)-r^{\dagger}E_-(-0))t\, , 
\nonumber \\ 
(F_-(+0)-r'^{\dagger}E_+(+0))t'=t^{\dagger}(F_-(-0)-E_-(-0)r)\, , 
\label{n3} \\
t^{\dagger}E_-(-0)t=E_-(+0) +r'^{\dagger}E_+(+0)r'- F_-(+0)r'- 
r'^{\dagger}F_+(+0) \, ,\nonumber \\  
t'^{\dagger}E_+(+0)t'=E_+(-0) +r^{\dagger}E_-(-0)r- F_+(-0)r- 
r^{\dagger}F_-(-0) \, . \nonumber 
\eea
``Rotating'' the amplitudes $E$ and $F$ by the unitary matrices $u_j$
we can bring all coefficients in eqs.\ (\ref{n3}) into the diagonal 
form in the transverse mode space. The equilibrium Green's functions 
$G_{1,2}$ deep inside first and second superconductors, which play the 
role of the source terms in equations defining the amplitudes, are 
proportional to unit matrix in this space. This means that the 
boundary conditions (\ref{n3}) become diagonal for the appropriately 
rotated amplitudes. In terms of the combinations of these amplitudes 
\bea 
Q^{\pm}_1=(u_4F_+(-0)u_1^{\dagger}\pm u_1F_-(-0)u_4^{\dagger}))/2 
\, , \;\;\;\;
Q^{\pm}_2=(u_3F_+(+0)u_2^{\dagger}\pm u_2F_-(+0)u_3^{\dagger}))/2 
\, ,  \nonumber \\
P^-_1=(u_4E_+(-0)u_4^{\dagger} -u_1E_-(-0)u_1^{\dagger})/2  
\, , \;\;\;\;
P^+_1=(u_4E_+(-0)u_4^{\dagger} +u_1E_-(-0)u_1^{\dagger})/2 -G_1   
\, ,  \nonumber \\
P^-_2=(u_3E_+(+0)u_3^{\dagger} -u_2E_-(+0)u_2^{\dagger})/2 
\, , \;\;\;\;
P^+_2=(u_3E_+(+0)u_3^{\dagger} +u_2E_-(+0)u_2^{\dagger})/2 -G_2 
\, , \nonumber 
\eea  
the boundary conditions (\ref{n3}) can be written as follows: 
\bea
P^-_1=P^-_2\, , \;\;\;\; Q^-_1=Q^-_2\, , \nonumber \\
Q^+_1-Q^+_2= \sqrt{R}(P^+_1+P^+_2+G_1+G_2)\, ,  \label{3} \\
G_1-G_2+P^+_1-P^+_2= \sqrt{R}(Q^+_1+Q^+_2) \, . \nonumber 
\eea 

Requirement that all deviations of the Green's functions from the 
equilibrium homogeneous values $G_{1,2}$, that are represented by 
the non-vanishing amplitudes $P$ and $Q$, decay away from the 
junction region (at $z\ra \pm \infty$) imposes another conditions 
on $P$ and $Q$. This condition has the same form as in the 
single-mode case \cite{b7}:  
\begin{equation}
G_jP^{\pm}_j=-P^{\pm}_jG_j=(-1)^{j+1} P^{\mp}_j\, , \;\;\;\;
G_jQ^{\pm}_j=Q^{\pm}_jG_j=(-1)^{j+1} Q^{\mp}_j\,, \;\;\;\; 
j= 1,2\, .
\label{4}\end{equation}
Combined, eqs.\ (\ref{3}) and (\ref{4}) give a system of 
linear matrix equations for the amplitudes $P$ and $Q$. Solving 
this system and using the normalization condition satisfied by 
the Green's functions $G_{1,2}\,$, $G_{1,2}^2=1$, \cite{b9} we 
get finally: 
\bea 
E_+=u_4^{\dagger}D(G_1+1)G_-G_+/Ku_4+G_1\, , \;\;\;\; 
F_+=u_4^{\dagger}\sqrt{R}(G_1+1)/Ku_1\, , \label{5} \\
E_-=u_1^{\dagger}D(G_1-1)G_-G_+/Ku_1+G_1 \, , \;\;\;\; 
F_-=u_1^{\dagger}\sqrt{R}(G_1-1)/Ku_4\, , \nonumber  
\eea 
where $G_{\pm}= (G_1\pm G_2)/2$ and $K= 1-DG_-^2$. The amplitudes 
$E$ and $F$ in eq.\ (\ref{5}) are given in the first electrode 
($z\ra -0$). Similar expressions can be obtained for the second 
electrode.

Equations (\ref{5}) solve the problem of electron transport in 
different types of junctions ranging from quantum point contacts 
with few propagating electron modes to short multi-mode SNS 
junctions with diffusive electron transport. They are valid for 
arbitrary ratio of the two values of the superconducting energy 
gap in the junctions electrodes, and can be applied, e.g.,  
to symmetric SNS junctions with the same gap and NS junctions 
with vanishing gap in one of the electrodes. The amplitudes 
(\ref{5}) define the Green's functions (\ref{2}) in the junction  
region and allow one to calculate different transport 
characteristics  
of the junction. For example, since the single-mode version of 
eqs.\ (\ref{5}) was the starting point of the theory~\cite{b11} 
of ac Josephson effect in high-transparency junctions, eqs.\ 
(\ref{5}) prove that this theory can be extended automatically 
to the multi-mode junctions. Another application of eqs.\ 
(\ref{5}) is presented in the next Section, where they are used 
to calculate equilibrium fluctuations of the supercurrent.

\section{Equilibrium supercurrent noise}

General expression for the current noise in terms of the 
quasiclassical Green's functions is derived in~\cite{b10}. 
In equilibrium, this equation reduces to the following 
result for the spectral density of current fluctuations~\cite{b1}: 
\begin{equation} 
S_I (\omega ) = \frac{e^2}{32\pi^2\hbar } \sum_{\pm \omega, \, 
\pm} \int d\epsilon f(\epsilon) (1-f(\epsilon \pm \omega))  
\mbox{Tr} [\rho_{\pm} (\epsilon) \sigma_z \rho_{\pm} (\epsilon 
\pm \omega) \sigma_z - \nu_{\pm} (\epsilon) \sigma_z \nu_{\mp} 
(\epsilon \pm  \omega )\sigma_z ] \, , 
\label{6} \ee  
where $\sigma_z$ is the Pauli matrix, the trace is taken over 
the transverse mode space and over electron-hole space, 
$\rho(\epsilon) \equiv E^R(\epsilon)-E^A(\epsilon)$, and 
$\nu(\epsilon) \equiv F^R(\epsilon)-F^A(\epsilon)$. The functions 
with superscripts $R,A$ denote retarded and advanced components 
of the Green's functions.  

In what follows, the noise (\ref{6}) is calculated for a symmetric 
junction between two identical superconductors which are close to 
``ideal'' BCS superconductors:
\be 
G_1^{R,A}(\varepsilon) = \frac{1}{((\epsilon\pm i\gamma_1)^2+ 
(\Delta \pm i\gamma_2)^2)} \left( \begin{array}{cc} 
\epsilon\pm i\gamma_1 & (\Delta \pm i\gamma_2)  e^{i
\varphi/2} \\ 
-(\Delta \pm i\gamma_2) e^{-i\varphi/2} & - (\epsilon\pm i\gamma_1) 
\end{array} \right) \, .  
\label{7} \ee 
Here $\varphi$ is the Josephson phase difference across the 
junction. Imaginary parts $\gamma_{1,2}$ of $\epsilon$, $\Delta$ 
arise from inelastic scattering, for instance the electron-phonon
scattering, and are assumed to be small, $\gamma_{1,2} \ll \epsilon\, 
,\Delta$. The functions $G_2^{R,A}$ in the second electrode are 
given by the same expression with $\varphi \ra -\varphi$. For 
superconductors described by the Green's functions (\ref{7}), the 
amplitudes (\ref{5}) contain resonant denominators $1-DG_-^2$, the 
contribution from which dominates the noise at low frequencies 
$\omega \ll \Delta/\hbar$. Indeed, in this case $G_-^2=
\tilde{\Delta}^2 \sin(\varphi/2)/(\tilde{\Delta}^2- 
\tilde{\epsilon}^2)$, where $\tilde{\Delta}= \Delta \pm i\gamma_2$ 
and $\tilde{\epsilon} =\epsilon\pm i\gamma_1$, and $1-DG_-^2$ nearly 
vanishes for $\varepsilon \simeq \varepsilon_k \equiv \pm 
\Delta [1-D_k\sin^2(\varphi/2)]^{1/2}$. Then, 
\be 
\rho(\epsilon),\nu(\epsilon) \propto \sum_{\pm} \frac{\gamma_k 
/2}{(\varepsilon \mp \varepsilon_k)^2+\gamma_k^2/4} \, ,
\label{8} \ee 
where $\gamma_k=2(\gamma_1(\varepsilon_k) \mp \gamma_2 
(\varepsilon_k) \varepsilon_k/\Delta)$. (Since $\gamma_1 
(-\varepsilon)= \gamma_1(\varepsilon)$  and 
$\gamma_2(-\varepsilon)=-\gamma_2(\varepsilon)$, the width 
$\gamma$ of the resonance is the same for both resonances in eq.\ 
(\ref{8}).) The two resonances correspond to the two discrete 
Andreev states per mode with energies $\pm\varepsilon_k$ in the 
subgap region. Their broadening $\gamma$ is caused by the 
inelastic transitions between these states and 
the continuum of states in the bulk electrodes. Expression for 
$\gamma$ in the case of electron-phonon transitions is given, 
e.g., in~\cite{b8}. 

Due to resonance (\ref{8}), we can take all 
non-resonant terms out of the integral over $\varepsilon$ in eq.\ 
(\ref{6}). Evaluating the non-resonant terms at $\varepsilon= 
\pm\varepsilon_k$ and intergrating the resonant denominators we 
obtain the spectral density of current fluctuations at low 
frequencies $\omega \sim \gamma$: 
\be
S_I(\omega) = \frac{1}{2\pi} \sum_{k=1}^N \left( \frac{I_k}{\cosh 
(\varepsilon_k/2T)} \right) ^2 \frac{\gamma_k }{ \omega^2+ 
\gamma^2_k } \, ,     
\label{9} \ee
where $I_k= (e\Delta^2/2\hbar) D_k \sin\varphi/ \varepsilon_k$ 
is the contribution to the supercurrent from one of the Andreev 
states in the $k$th mode. 

Equation (\ref{9}) determines the equilibrium noise of the 
supercurrent in short Josephson junctions. It has a simple 
interpretation in terms of the fluctuations of the occupation 
factors of Andreev states carrying the supercurrent. All states 
are occupied independently one from another. If a state with 
energy $\pm\varepsilon_k$ is occupied, it gives contribution 
$\pm I_k$ to the supercurrent; if it is empty, the contribution 
to the supercurrent is zero. These two situations are realized 
with the probabilities $f(\pm \varepsilon_k)$ and $1-f(\pm 
\varepsilon_k)$, respectively, and the transitions between them 
occur with the characteristic rate $\gamma$. Calculation of the 
spectral density of this simple classical stochastic process 
reproduces eq.\ (\ref{9}). This equation was conjectured 
in~\cite{b1} on the basis of the similar interpretation of the 
noise in the single-mode ballistic junctions. The calculation 
presented in this work provides its rigorous proof. 

At temperatures close to the superconducting critical 
temperature $T_c$ of the junction electrodes, the energy gap is 
small, $T\gg \Delta$, and the transition rates $\gamma_k$ and 
the occupation probabilities of Andreev states are independent 
of the state energy, and eq.\ (\ref{9}) can be 
simplified further. In this high-temperature regime, one also has 
to take into account the non-resonant contribution to the current 
noise from the continuum part of the junction spectrum which can 
become comparable to the resonant contribution (\ref{9}). As can 
be seen  from eq.\ (\ref{6}), this non-resonant contribution 
coincides at $T\gg \Delta$ with the regular equilibrium thermal 
noise of a normal junction. Combining the two contributions we 
obtain the noise in the high-temperature limit:  
\be
S_I(\omega) = \frac{1}{2\pi} \left( \sum_{k=1}^N I_k^2 \right) 
\frac{\gamma }{ \omega^2+ \gamma^2 } +S_N\, ,      
\label{10} \ee 
where $S_N=GT/\pi$ and $G=(e^2/\pi \hbar) \sum_{k=1}^{N} D_k$ is 
the junction conductance. 

Equation (\ref{10}) can be evaluated if the junction has known 
distribution of transparencies. An important example is provided 
by the regular multi-mode disordered SNS junction with diffusive 
electron transport. This type of junctions is characterized by 
the quasicontinuous  Dorokhov's distribution of transparencies, 
$\sum_k = (\pi \hbar G/2e^2) \int_0^1 dD/D(1-D)^{1/2}$. Calculating 
the integral over $D$ we find spectral density of current noise 
at large temperatures: 
\be   
S_I(\omega) = \frac{G\Delta^2}{2\hbar }\cos^2 \frac{\varphi}{2} 
\left(\frac{\varphi}{\sin \varphi}-1 \right) \frac{\gamma 
}{ \omega^2+ \gamma^2 } +S_N \, , \;\;\;\;  
\varphi \in [-\pi,\pi] \, .    
\label{11} \ee 
The phase dependence of the supercurrent-noise part of eq.\ 
(\ref{11}) is plotted in Fig.\ 1. Equation (\ref{11}) shows that 
if $T\ll \Delta^2/\hbar \gamma $, this noise gives the dominant 
contribution to the current noise at frequencies $\omega \sim 
\gamma$. The noise can 
also be larger than the average supercurrent $I(\varphi)= (\pi 
G\Delta^2 /4eT) \sin \varphi$. If $\Delta/T<(e^2/\hbar G)^{1/2}$, 
typical r.m.s. of the supercurrent noise (in the frequency range 
$\gamma$) is larger than $I(\varphi)$. This means that the 
supercurrent noise should qualitatively change the behavior of the 
SNS junctions at temperatures close to $T_c$ and should also be 
observable at low temperatures.   

\begin{figure}[htb]
\setlength{\unitlength}{1.0in}
\begin{picture}(3.2,2.45) 
\put(1.4,-0.15){\epsfxsize=3.2in\epsfysize=2.5in\epsfbox{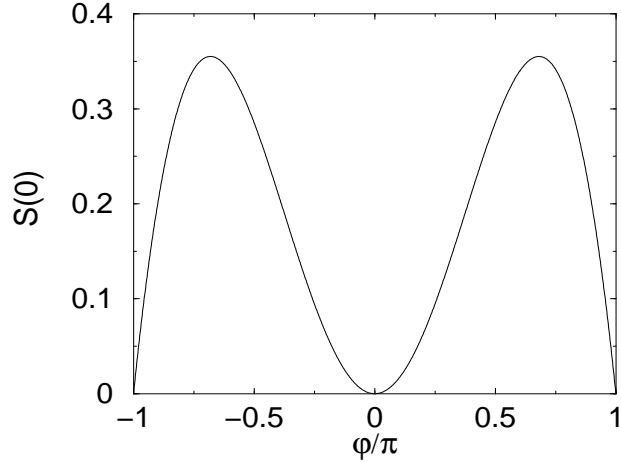}}
\end{picture}
\caption{Dependence of the zero-frequency spectral density 
$S(0)=(S_I(0)-S_N)/(G\Delta^2/2\hbar \gamma)$ of the supercurrent 
noise in short disordered SNS junctions on the Josephson phase 
difference $\varphi$ at large temperatures $T\gg \Delta$.   
All quantities are defined as in the eq.\ \protect (\ref{11}). 
\label{fig:1}}
\end{figure}

\section{Conclusion} 

In conclusion, the equilibrium supercurrent noise has been studied 
in short Josephson junctions with arbitrary electron transmission 
properties. Characteristic qualitative features of the noise are 
its dependence on the Josephson phase difference across the 
junction, and the cutoff frequency determined by the relaxation 
rate, which can be much smaller than the temperature and 
superconducting energy gap. The r.m.s. of the noise is larger 
that the average supercurrent at temperatures close to the 
critical temperature of the junction electrodes. A formal result 
of this work is the general expression for the quasiclassical 
Green's functions of the multi-mode junctions that can be used to 
study other transport characteristics, for instance, the 
non-equilibrium noise in the MAR regime.

\section*{Acknowledgments}

This work was supported in part by ONR grant \# N00014-95-1-0762 
and by the U.S. Civilian Research and Development Foundation under 
Award No. RP1-165.

\end{document}